\documentclass[twocolumn,showpacs]{revtex4}
\usepackage{amssymb}
\usepackage{eepic}
\usepackage{graphicx}

\begin{document}

\title{Real-time wavelet-transform spectrum analyzer
for the investigation of $1/f^{\alpha}$ noise}

\author{Doriano Brogioli}
\author{Alberto Vailati}
\affiliation{Dipartimento di Fisica and Istituto Nazionale per la Fisica 
della Materia, Universit\`a di Milano, via Celoria 16,
20133 Milano, Italy}

\begin{abstract}
A wavelet transform spectrum analyzer operating in real time within the
frequency range $3\times 10^{-5} - 1.3\times10^5 \mathrm{Hz}$
has been implemented on a low-cost Digital Signal
Processing board operating at $150\mathrm{MHz}$.
The wavelet decomposition of the
signal allows to efficiently process non-stationary signals dominated by
large amplitude events fairly well localized in time, thus providing the
natural tool to analyze processes characterized by
$1/f^{\alpha}$ power spectrum. The
parallel architecture of the DSP allows the real-time processing of the
wavelet transform of the signal sampled at $0.3\mathrm{MHz}$.
The bandwidth is about $220\mathrm{dB}$, almost ten decades.
The power spectrum of the scattered intensity is
processed in real time from the mean square value of the wavelet
coefficients within each frequency band. The performances of the spectrum
analyzer have been investigated by performing Dynamic Light
Scattering experiments on colloidal suspensions and by comparing the
measured spectra with the correlation functions data obtained with a
traditional multi tau correlator. In order to asses the potentialities of
the spectrum analyzer in the investigation of processes involving
a wide range of timescales,
we have performed measurements on a model system where fluctuations in the
scattered intensities are generated by the number fluctuations
in a dilute colloidal suspension illuminated by a wide beam. This system
is characterized by a power-law spectrum with exponent $-3/2$ in
the scattered intensity fluctuations.
The spectrum analyzer allows to recover the power spectrum with a dynamic 
range spanning about $8$ decades.
The advantages of wavelet analysis versus correlation analysis in the 
investigation of processes characterized by a wide distribution of time
scales and non-stationary processes are briefly discussed.
\end{abstract}
\maketitle
\newpage

\section{Introduction}

The investigation of random processes characterized by a wide range of
timescales is becoming increasingly important in many experimental fields
ranging from the investigation of earthquakes in earth-science, to turbulence
in fluids, or the price fluctuations in financial markets 
\cite{rev_1of_noise,sornette,sethna,mantegna}.
In particular, many random processes exhibit a scale
invariant structure. The self-similarity of the signal is reflected in the
power law behavior of the power spectrum, which lacks characteristic time
scales.

Traditional spectral decomposition techniques often 
fail when applied to a self
similar signal. This is due to the fact that such a signal is 
typically the
superposition of bursts occurring at many different timescales. Although
rare, long bursts having a very large amplitude provide much of the energy
content of the signal, while short bursts with small amplitude, although
very frequent, give small contribution to the energy of the signal. The
energy is therefore fairly well localized in time and the signal is
non-stationary. This localization prevents Fourier Analisys to work as an
effective tool for this kind of signal.

However,
a scale invariant signal can be efficiently analyzed by using the
renormalization group approach devised in statistical physics to describe
systems at a critical point \cite{lebellac}.
The self similar system goes through a consecutive sequence of coarse
grainings which generate a new signal with statistical properties similar to
those of the original one.

In practice the renormalization group analisys of a signal is achieved by
using the wavelet transform technique. Basically the signal is 
convolved with a chosen mother wavelet function. The mother
wavelet consists in a truncated wave packet with central frequency $\omega$
and bandwidth $\Gamma$, localized at time $t$, with length
$l\propto 1/\Gamma$. The mean square amplitude of this
component provides the power spectral amplitude of the signal in the
frequency range $\left(\omega -\Gamma ,\omega +\Gamma\right)$. 
An in-quadrature component of the signal
is also simultaneously processed to recover the discarded lower frequency
component, representing the coarse grained
signal. The mother wavelet is
then rescaled on a larger scale and the process iterated.

In this paper we describe a real-time wavelet transform spectrum analyzer
(WTSA) working in the frequency range 
$3\times 10^{-5} - 1.3\times 10^{5} \mathrm{Hz}$.
The spectrum analyzer is built around a Texas Instruments C6711 DSP
Starter Kit (DSK). This board incorporates a C6711 DSP running at 
$150\mathrm{MHz}$
together with a parallel port to communicate with a Personal computer and
various connectors to interface the board to the real world. Solutions
based on wavelet analysis by means of the post processing on a dedicated
personal computer of the digitized signal have recently been proposed
\cite{wegdam,sprik}. The highly parallel processing architecture of the DSP
allows the real-time processing of the wavelet transform of the signal
sampled at $0.3\mathrm{MHz}$. The bandwidth spans almost ten decades.

We will present applications to Dynamic Light Scattering (DLS) from
colloidal suspensions and compare the results obtained with the WTSA with
the correlation function obtained with a traditional multi tau correlator.
Dynamic Light Scattering has been used very extensively to investigate 
stochastic fluctuations
in the intensity of the light scattered by suspended particles undergoing
brownian motion, aggregation processes, gelation processes, critical
dynamics \cite{pecora1, pecora2, brown}. In recent times, processes
characterized by a wide distribution of relaxation times, often leading to
stretched-exponential decay of the autocorrelation function, have attracted
much interest (see Refs. \cite{cipelletti2000,cipelletti2001} and 
references therein). The development of
log-scale correlators has allowed the investigation of processes extending
from about $100\mathrm{ns}$ up to hours. However, large time lags require
very long acquisitions times to get good statistical accuracy in the
correlation tails.

The wavelet analisys of the fluctuations in the intensity of the scattered
light provides higher accuracy in shorter times when compared to DLS when
non-stationary self similar signals are processed. This is due to the fact
that the localized wavelets used to process the signal are very efficient in
isolating the well localized features of a self-similar signal, and this
provides a shorter convergence time of the processing to get results with a
chosen statistical accuracy.

To asses the potentialities of the WTSA in investigating time-scale
invariant systems we have devised a simple model system giving rise to a
power-law spectrum with exponent $-3/2$. This system is made up by strongly
diluted brownian particles diffusing within a large sampling volume. We show
that the WTSA is able to measure the power spectrum within a dynamic range
as large as 8 decades. We also show that the WTSA requires a measurement
time roughly $100$ times smaller than that needed by a correlator
to get the same statistical accuracy at large lag times.

\section{Wavelet analysis}
\label{sezione_wavelet_analysis}

The wavelet analisys of signals is becoming increasingly popular within the
scientific community. With respect to traditional Fourier decomposition
techniques, wavelet analisys allows to characterize signals having a large
bandwidth by using a small number of coefficients.

Basically, instead of decomposing the signals into armonic oscillations as
in Fourier analisys, the signal gets decomposed into the superposition of
wave packets (WP). The WP are localized both in time and in frequency. These
wave packets are obtained by rescaling a chosen Mother Wavelet (MW) 
$\Psi\left(t\right)$ with a proper scale $s$ so as to obtain a wavelet
function $\varphi_{t_0,s}\left(t\right)=\Psi\left(\frac{t-t_0}{s}\right)$.
The wavelets exhibit the same functional form of the MW, but are centered 
around time $t_0$ and their
bandwidth is a factor $s$ larger than that of the MW. Therefore, if the
signal has a bump at time $t$ with a certain duration $\Delta t$ the wavelet
decomposition will likely give rise to a wave packet localized at $t$ with a
scale $s$ proportional
to the duration of the bump. This allows a very efficient
decomposition of spiky signals and non-stationary signals in general. In
practice the wavelet analysis is usually performed by using the popular
Discrete Wavelet Transform (DWT) algorithm \cite{numerical_recipes}. To apply
this algorithm, one starts with a sampled discrete signal $y_n^0$ made
up by $2^N$ samplings, where $N$ is an integer. The signal undergoes
successive coarse grainings and band-pass filterings by convolution with a
chosen wavelet set. The scale $s$ of the wavelets used during the coarse
grainings changes by a factor $2$ at each step, so that the scaling of the
signal at the $k$-th step is given by $s_k=s_0 2^k$. On the contrary,
the number of samplings in the coarse grained signal decreases as $2^N/2^k$.
Therefore, as the scale increases due to the successive scale doublings,
the resolution in the temporal localization of the wavelet decreases
accordingly. In the following we will give a brief and more rigorous
overview on how the power spectrum is processed from the sampled signal.

Consider a discrete signal $y_n^0$ sampled at a given sample
frequency $f_s$.
According to Nyquist theorem the higher angular frequency it contains 
is $\omega_0=\pi f_s$, that corresponds to the function 
$y_n^0=\left(-1\right)^n$. The original signal $y_n^0$ is passed
through a high-pass filter, giving the wavelet coefficients $c^0_n$,
representing the high frequency details of the signal in the range
$\left(\omega_0/2,\omega_0\right)$,
and through a low-pass filter, giving a coarse grained signal $y^1_n$.
Both the filters are implemented by convolving the signal with 
suitable filter coefficients $g_n$ (the MW) and $h_n$; they are
localized in $n$ so that the only non vanishing elements are the ones with
$0 \le n < M$: 
\begin{equation}
\left\{ 
\begin{array}{l}
c^0_j=\sum_n{y^0_n g_{n-2j}} \\
y^1_j=\sum_n{y^0_n h_{n-2j}}
\end{array}
\right. .
\end{equation}
Both the filtered signals $c^0_n$ and $y^1_n$ are sampled at half
the original frequency $f_s$. The two filter coefficients $h_n$ and $g_n$
are such that the coarse grained signal $y_n^1$, along with the
details $c^0_n$, 
allow to recover the whole original signal $y^0_n$: they are referred to as 
quadrature mirror filters. In the measurements we performed we used Daubechies
wavelets \cite{numerical_recipes}.

The procedure is then iterated, using $y^1_n$ as the new signal;
a high frequency component $c_n^1$ is obtained together with a 
coarse grained signal $y^2_n$, and so on. In general, the following
recursion formulas apply:
\begin{equation}
\left\{ 
\begin{array}{l}
c^{k}_j=\sum_n{y^k_n g_{n-2j}} \\
y^{k+1}_j=\sum_n{y^k_n h_{n-2j}}
\end{array}
\right. .
\end{equation}
At step $k$, the signal $y^k_n$ is coarse grained to obtain 
$y^{k+1}_n$; the high frequency details, spanning the angular frequency range
$\left(\omega_0/2^{k+1},\omega_0/2^k\right)$ are the wavelet
coefficients $c^k_n$.

The power spectrum of the signal can then be derived from the mean square
average of the details
$\left<\left| c_n^k\right|^2\right>_n$.
 
With the
traditional Fourier transform, the analyzed wavelengths are discretized, due
to the finiteness of the sample; the spacing between two consecutive
wavenumbers is constant. Unlikely, with wavelet analysis, the spacing is
exponential: the mean square amplitude of the wavelet coefficients
represents the power contained in an octave. 

The exponential spacing is particularly suited to measure phenomena
spanning many decades in wavenumber.
This feature is analogous to the multi tau analysis done by the best
commercial correlators. Though the power spectrum and the correlation
function are connected by the well known Wiener-Kintchine theorem, in
many actual cases it is not possible to recover the power spectrum by the
measurement of the correlation function. In general,
the exponential spaced correlation function given by a multi-tau
correlator cannot be safely Fourier transformed into the power spectrum,
over many decades. Moreover, in some cases
Wiener-Kintchine theorem does not apply. This is the case of some scale
invariant and non-stationary signals; in Sect. \ref{sistema_modello} we present
an example in which the power spectrum obtained by the wavelet analysis
correctly shows a power law behaviour, while a multi tau correlator
seemingly says that the signal is delta correlated.

The analogy between the wavelet analysis and the multi tau correlation also
includes the possibility of performing the required processing
in real time, as the sample is
acquired. With Fourier analysis, as the sample becomes larger, we must
analyze more and more wavelengths, and this requires that all the sample is
recorded and re-analyzed. On the contrary, as time goes on, the multi
tau correlator averages the older samples; as a new, longer delay time
is available, the correlation function is evaluated from only a few registers.
This allows to extend the dynamic range in time delay over many decades,
without the need to record the samples with the same, huge dynamic range.
The same thing happens with the WTSA. The iterative procedure is
implemented by using $N$, almost identical levels $k$. 
The memory needed scales as $N$ while the time
needed by the processing does not change at all, as we will show in Sect.
\ref{sec_programma}; on the other hand, the frequency range scales as $2^N$.

\section{The Wavelet Transform Spectrum Analyzer}

The spectrum analyzer is built around a Texas Instruments C6711 DSP Starter
Kit (DSK), graciously provided by Texas Instruments DSP University Program.

This board incorporates a C6711 DSP running at $150\mathrm{MHz}$,
together with a parallel port to
communicate with a Personal computer and various connectors to interface the
board to the real world. The software can be developed on a Personal
Computer by using TI Code Composer Studio and then downloaded on the board.
The board and the bundled software are very cheap compared to a real time
correlator board, the purchase price being of the order of 300\$ for
educational institutions.

The core of the board is made up by a Digital Signal Processor, a
single-chip programmable device including a CPU specifically suited for
digital signal processing
\cite{manuale_DSP_CPU}. The DSP is a TMS320C6711, built by Texas
Instruments, one of the fastest processors currently available. It is based
on VelociTI, a high-performance, advanced very-long-instruction-word
architecture, and its CPU can perform operations on double precision ($64$
bit) floating point numbers.

The CPU is divided into two almost identical sides; each of them includes $16$
registers, one word ($32$ bit) long. Couples of registers can be used to
represent one floating point, double precision register. Each section
includes four ALUs, operating simultaneously; each ALU can perform only a
given set of operations. Data paths from registers to memory and from a
register in one side to an ALU in the other are also provided.

The CPU works at $150\mathrm{MHz}$.
At each clock cycle, up to eight instructions are
fetched, dispatched and decoded, that is, up to one instruction for each
ALU. All the fixed point instructions are executed in one clock cycle; this
means that the CPU can perform $1.2\times 10^{9}$ instructions per second, a
figure comparable to the number of MFLOPS available on a $2\mathrm{GHz}$
Pentium IV processor. Some floating point instructions require more than 
one clock
cycle to return the results (execution latency); generally, the same unit
can start a new instruction after a few clock cycles (functional unit
latency), before the results of the previous operations are availabe. For
example, the functional unit latency of a double precision floating point
multiplication is $4$ clock cycles, and the execution latency is $9$ clock
cycles. Consider, for example, an ALU starting a multiplication at
$t=0$: at $t=4$ clock cycles it can start a new multiplication and
at $t=9$ clock cycles the
results of the first multiplication are available on the registers.
Since there are two ALUs
able to perform floating point multiplications, the CPU can do $60\times
10^{6}$ multiplications per second, and, simultaneously, much more floating
point additions, using other two ALUs. This structure requires a
sophisticated, parallel programming technique, resulting in extremely fast,
real time programs.

The DSP includes a two level memory \cite{manuale_DSP_perif}.
The first level is composed by two
$4$Kbytes cache memories, L1P for the program instructions and L1D for the
data. The second level, L2, is a $64$Kbytes RAM, that can be configured
partially as a SRAM or as a chache for an external memory. Since our program
and data fit in the SRAM, we didn't use the slower external RAM and we
configured all the L2 as SRAM. Reads and writes in the L1P and L1D memories
are performed in four clock cycles; obviously, L2 is slower and a miss in
the L1 cache results in a CPU stall.

The DSP includes many other peripherals; among them, two counter-timers,
TIMER1 and TIMER2 \cite{manuale_DSP_perif}.
They are broadly configurable through a configuration
register. The input can be connected to the system clock or to an external
pin, driven by a CMOS signal, thus selecting if the device is acting as a
counter of external events or as a timer. The count register can be read and
used by the program. Based on the counts and the value of a period register,
an output is driven, in order to generate square waves of different duty
cycle. Moreover, the output can drive an interrupt. We used TIMER1 as a
counter of external events. In this way, the input of the spectrum analyzer
can be driven by a discrete pulse train, as in the application to 
DLS discussed below, where the timer is connected to the output
of a photon counting photomultiplier. Alternatively, by using a voltage to
frequency converter, the input of the timer could be driven by an analog
signal, such as the one from a photodiode. The other timer, TIMER2,
generates a periodic interrupt, in order to define the integration period:
the interrupt service routine reads the counts of TIMER1 and resets it,
passing the number of counts to the processing program.

Texas Instruments provides a DSK (DSP Starter Kit), including a board with
the DSP and some peripherals; among them, a device that allows to connect
the board to a PC through a parallel port. The PC can read and write the
memory of the DSP, without halting the CPU, and can run a program. The PC
uploads the program that evaluates the power spectrum of the signal in the
SRAM, then runs it, and periodically reads the results.

The software consists of two parts, one for the DSP and one for the host
computer. Texas Instruments provides Code Composer Studio (CCS), that
includes a C compiler and an assembler for the DSP. The program for the DSP
has been written in assembler code, since in this way a high processing
speed is obtained in conjunction with an accurate timing. Moreover, the
provided software includes some libraries for building host programs that
interact with the DSP, that is, to upload a program, run it, and read some
memory locations in the DSP \cite{manuale_DSP_progr}.
All the software we developed is freely available. \cite{WTSA_URL}

\section{Discrete wavelet transform program}
\label{sec_programma}

With respect to the hardware correlators traditionally used within the
Dynamic Light Scattering community, the use of a DSP allows the flexible
implementation of many processing algorithms simply by dowloading a
different processing code onto the board. These algorithms include FIR and IIR
filters, cross correlation, convolution filters etc.

In this section, we discuss the implementation of the discrete, real time
wavelet algorithm on the DSK board.

The overall data structure is shown in Fig. 1, along with a simplified
description of the data flow. The higest sampling frequency is
about $0.3\mathrm{MHz}$ and the lowest frequency corresponds to a period of
some hours.
\begin{figure*}
\begin{picture}(480,200)(0,0)
\dottedline{3}(110.000000,0.000000)(110.000000,200.000000)
\path(120.000000,120.000000)(120.000000,100.000000)(150.000000,100.000000)(150.000000,120.000000)
\path(120.000000,140.000000)(120.000000,120.000000)(150.000000,120.000000)(150.000000,140.000000)
\path(120.000000,160.000000)(120.000000,140.000000)(150.000000,140.000000)(150.000000,160.000000)
\path(120.000000,180.000000)(120.000000,160.000000)(150.000000,160.000000)(150.000000,180.000000)(120.000000,180.000000)
\put(125.000000,107.500000){$y^0_{n-3}$}
\put(125.000000,127.500000){$y^0_{n-2}$}
\put(125.000000,147.500000){$y^0_{n-1}$}
\put(125.000000,167.500000){$y^0_{n}$}
\path(221.000000,150.000000)(221.000000,180.000000)(246.000000,165.000000)(221.000000,150.000000)
\put(226.000000,162.500000){$\times$}\path(233.500000,157.500000)(234.500000,161.500000)(238.500000,160.500000)
\path(221.000000,100.000000)(221.000000,130.000000)(246.000000,115.000000)(221.000000,100.000000)
\put(226.000000,112.500000){$\times$}\path(233.500000,107.500000)(234.500000,111.500000)(238.500000,110.500000)
\path(155.000000,180.000000)(170.000000,176.000000)(221.000000,176.000000)
\path(155.000000,100.000000)(170.000000,174.000000)(184.000000,174.000000)(184.000000,124.000000)(221.000000,124.000000)
\path(221.000000,174.000000)(186.000000,174.000000)(186.000000,126.000000)(221.000000,126.000000)
\path(201.000000,160.000000)(211.000000,156.000000)(221.000000,156.000000)
\path(201.000000,150.000000)(211.000000,154.000000)(221.000000,154.000000)
\path(201.000000,110.000000)(211.000000,106.000000)(221.000000,106.000000)
\path(201.000000,100.000000)(211.000000,104.000000)(221.000000,104.000000)
\put(191.000000,102.500000){$g_j$}
\put(191.000000,152.500000){$h_j$}
\path(120.000000,70.000000)(160.000000,70.000000)(160.000000,90.000000)(120.000000,90.000000)(120.000000,70.000000)
\put(125.000000,77.500000){Status}
\put(161.000000,83.500000){=2?}
\put(112.000000,83.500000){+}
\path(120.000000,40.000000)(167.000000,40.000000)(167.000000,60.000000)(120.000000,60.000000)(120.000000,40.000000)
\put(125.000000,47.500000){Counter}
\path(120.000000,10.000000)(165.000000,10.000000)(165.000000,30.000000)(120.000000,30.000000)(120.000000,10.000000)
\put(125.000000,17.500000){$\sum_j\left|c_j^0\right|^2$}
\path(235.000000,5.000000)(235.000000,35.000000)(210.000000,20.000000)(235.000000,5.000000)
\put(222.000000,17.500000){$\cdot^2$}\put(160.000000,190.000000){Level 0}
\path(160.000000,80.000000)(270.000000,80.000000)
\path(187.000000,83.000000)(190.000000,80.000000)(187.000000,77.000000)
\path(246.000000,165.000000)(266.000000,165.000000)(266.000000,170.000000)(270.000000,170.000000)
\path(253.000000,168.000000)(256.000000,165.000000)(253.000000,162.000000)
\path(120.000000,170.000000)(110.000000,170.000000)
\path(120.000000,80.000000)(110.000000,80.000000)
\path(246.000000,115.000000)(253.000000,115.000000)(253.000000,20.000000)(235.000000,20.000000)
\path(250.000000,63.000000)(253.000000,60.000000)(256.000000,63.000000)
\path(260.000000,80.000000)(260.000000,150.000000)(236.000000,150.000000)(236.000000,159.000000)
\put(260.000000,80.000000){\circle*{2}}
\path(260.000000,100.000000)(236.000000,100.000000)(236.000000,109.000000)
\put(260.000000,100.000000){\circle*{2}}
\path(205.000000,80.000000)(205.000000,50.000000)(167.000000,50.000000)
\path(189.000000,53.000000)(186.000000,50.000000)(189.000000,47.000000)
\put(205.000000,80.000000){\circle*{2}}
\path(210.000000,20.000000)(165.000000,20.000000)
\path(190.500000,23.000000)(187.500000,20.000000)(190.500000,17.000000)
\put(250.000000,175.000000){$y^1_m$}
\put(247.000000,120.000000){$c^0_m$}
\dottedline{3}(270.000000,0.000000)(270.000000,200.000000)
\path(280.000000,120.000000)(280.000000,100.000000)(310.000000,100.000000)(310.000000,120.000000)
\path(280.000000,140.000000)(280.000000,120.000000)(310.000000,120.000000)(310.000000,140.000000)
\path(280.000000,160.000000)(280.000000,140.000000)(310.000000,140.000000)(310.000000,160.000000)
\path(280.000000,180.000000)(280.000000,160.000000)(310.000000,160.000000)(310.000000,180.000000)(280.000000,180.000000)
\put(285.000000,107.500000){$y^1_{m-3}$}
\put(285.000000,127.500000){$y^1_{m-2}$}
\put(285.000000,147.500000){$y^1_{m-1}$}
\put(285.000000,167.500000){$y^1_{m}$}
\path(381.000000,150.000000)(381.000000,180.000000)(406.000000,165.000000)(381.000000,150.000000)
\put(386.000000,162.500000){$\times$}\path(393.500000,157.500000)(394.500000,161.500000)(398.500000,160.500000)
\path(381.000000,100.000000)(381.000000,130.000000)(406.000000,115.000000)(381.000000,100.000000)
\put(386.000000,112.500000){$\times$}\path(393.500000,107.500000)(394.500000,111.500000)(398.500000,110.500000)
\path(315.000000,180.000000)(330.000000,176.000000)(381.000000,176.000000)
\path(315.000000,100.000000)(330.000000,174.000000)(344.000000,174.000000)(344.000000,124.000000)(381.000000,124.000000)
\path(381.000000,174.000000)(346.000000,174.000000)(346.000000,126.000000)(381.000000,126.000000)
\path(361.000000,160.000000)(371.000000,156.000000)(381.000000,156.000000)
\path(361.000000,150.000000)(371.000000,154.000000)(381.000000,154.000000)
\path(361.000000,110.000000)(371.000000,106.000000)(381.000000,106.000000)
\path(361.000000,100.000000)(371.000000,104.000000)(381.000000,104.000000)
\put(351.000000,102.500000){$g_j$}
\put(351.000000,152.500000){$h_j$}
\path(280.000000,70.000000)(320.000000,70.000000)(320.000000,90.000000)(280.000000,90.000000)(280.000000,70.000000)
\put(285.000000,77.500000){Status}
\put(321.000000,83.500000){=2?}
\put(272.000000,83.500000){+}
\path(280.000000,40.000000)(327.000000,40.000000)(327.000000,60.000000)(280.000000,60.000000)(280.000000,40.000000)
\put(285.000000,47.500000){Counter}
\path(280.000000,10.000000)(325.000000,10.000000)(325.000000,30.000000)(280.000000,30.000000)(280.000000,10.000000)
\put(285.000000,17.500000){$\sum_j\left|c_j^1\right|^2$}
\path(395.000000,5.000000)(395.000000,35.000000)(370.000000,20.000000)(395.000000,5.000000)
\put(382.000000,17.500000){$\cdot^2$}\put(320.000000,190.000000){Level 1}
\path(320.000000,80.000000)(430.000000,80.000000)
\path(347.000000,83.000000)(350.000000,80.000000)(347.000000,77.000000)
\path(406.000000,165.000000)(426.000000,165.000000)(426.000000,170.000000)(430.000000,170.000000)
\path(413.000000,168.000000)(416.000000,165.000000)(413.000000,162.000000)
\path(280.000000,170.000000)(270.000000,170.000000)
\path(280.000000,80.000000)(270.000000,80.000000)
\path(406.000000,115.000000)(413.000000,115.000000)(413.000000,20.000000)(395.000000,20.000000)
\path(410.000000,63.000000)(413.000000,60.000000)(416.000000,63.000000)
\path(420.000000,80.000000)(420.000000,150.000000)(396.000000,150.000000)(396.000000,159.000000)
\put(420.000000,80.000000){\circle*{2}}
\path(420.000000,100.000000)(396.000000,100.000000)(396.000000,109.000000)
\put(420.000000,100.000000){\circle*{2}}
\path(365.000000,80.000000)(365.000000,50.000000)(327.000000,50.000000)
\path(349.000000,53.000000)(346.000000,50.000000)(349.000000,47.000000)
\put(365.000000,80.000000){\circle*{2}}
\path(370.000000,20.000000)(325.000000,20.000000)
\path(350.500000,23.000000)(347.500000,20.000000)(350.500000,17.000000)
\put(410.000000,175.000000){$y^2_l$}
\put(407.000000,120.000000){$c^1_l$}
\dottedline{3}(430.000000,0.000000)(430.000000,200.000000)
\put(440.000000,190.000000){Level 2}
\path(440.000000,170.000000)(430.000000,170.000000)
\dottedline{2}(440.000000,170.000000)(470.000000,170.000000)
\path(440.000000,80.000000)(430.000000,80.000000)
\dottedline{2}(440.000000,80.000000)(470.000000,80.000000)
\path(45.000000,30.000000)(95.000000,30.000000)(95.000000,50.000000)(45.000000,50.000000)(45.000000,30.000000)
\put(50.000000,37.500000){TIMER2}
\path(0.000000,120.000000)(50.000000,120.000000)(50.000000,140.000000)(0.000000,140.000000)(0.000000,120.000000)
\put(5.000000,127.500000){TIMER1}
\path(56.000000,160.000000)(84.000000,160.000000)(84.000000,180.000000)(56.000000,180.000000)(56.000000,160.000000)
\put(61.000000,167.500000){ISR}
\path(67.000000,160.000000)(70.000000,163.000000)(73.000000,160.000000)
\path(84.000000,170.000000)(110.000000,170.000000)
\path(94.000000,173.000000)(97.000000,170.000000)(94.000000,167.000000)
\path(70.000000,50.000000)(70.000000,160.000000)
\path(70.000000,80.000000)(110.000000,80.000000)
\put(70.000000,80.000000){\circle*{2}}
\path(67.000000,62.000000)(70.000000,65.000000)(73.000000,62.000000)
\path(87.000000,83.000000)(90.000000,80.000000)(87.000000,77.000000)
\path(25.000000,140.000000)(25.000000,170.000000)(56.000000,170.000000)
\path(22.000000,152.000000)(25.000000,155.000000)(28.000000,152.000000)
\put(10.000000,85.000000){Input}
\path(25.000000,120.000000)(25.000000,95.000000)
\path(22.000000,104.500000)(25.000000,107.500000)(28.000000,104.500000)
\end{picture}
\caption{Data structure and data flow of the wavelet transform spectrum analyzer.}
\end{figure*}
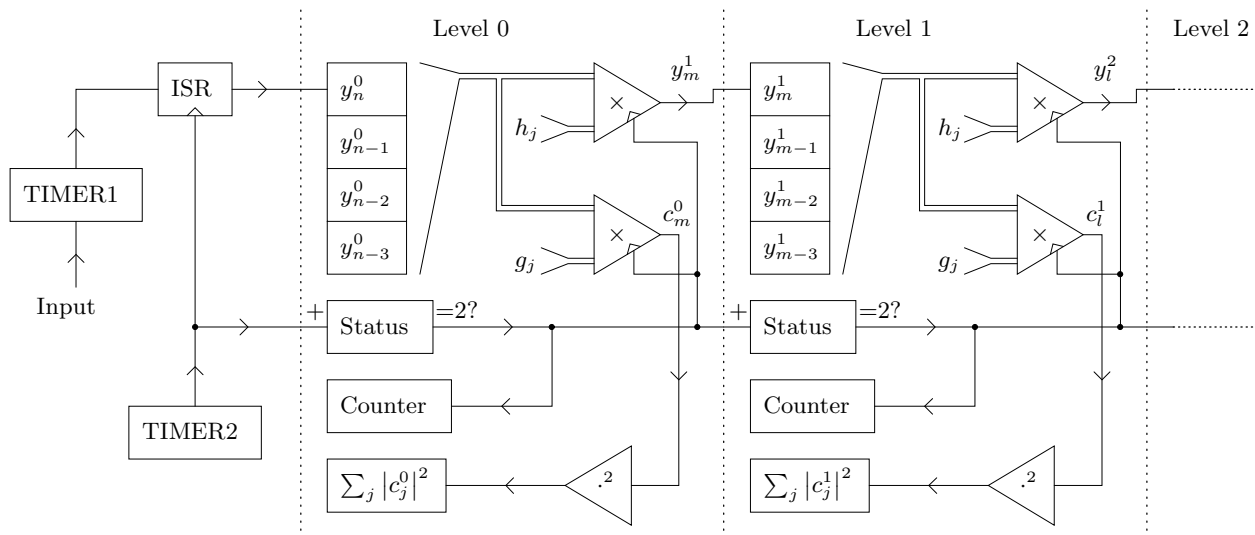
Every frequency octave represents a level in the wavelet processing
algorithm. Figure 1 shows only levels 0 and 1 and part of level 2. The
entire structure is made up by $N=32$ levels, each corresponding to a
measured frequency. Every level contains a circular queue $y_j^k$ of
double precision floating points to store the last $M$ values of the coarse
grained signal, where $M$ is the length of the MW. The $N$
queues are managed by using the circular addressing mode of the DSP; this
allows only values of $M$ that are powers of two. Therefore, the Daubechies
wavelets our program can use are those with length $M=4$, $8$, $16$, 
$32$ and $64$. We
will discuss the implementation of the Daub4 $M=4$ transform.

Level 0 is driven by TIMER1, configured as a counter, which counts the
digitized pulses at the input of the spectrum analyzer within a selected
bin. The integration time is defined by TIMER2, configured as a timer, so
that its output signal generates an interrupt at 
$f_s=375\mathrm{KHz}$, corresponding to the sampling frequency. At
each sampling time the number of pulses counted by TIMER1 is fed to the
first element of the shift register $y_n^0$ in Level 0 and the other
cells are shifted down. Every two sample times the content of the
shift register is convolved with the two wavelet coefficient banks $g_n$ 
and $h_n$, the MW and the smoothing filter coefficients respectively.
The result of the convolution with the bank $h_n$ feeds the shift
register $y_n^1$ in Level 1 and gives rise to a coarse grained replica
of the signal at half the sampling frequency. The in-quadrature component
obtained from the convolution with bank $g_n$ contains the wavelet
coefficients $c_n^0$ which specify the local amplitude of the band-pass
filtered signal in the frequency range $\left(\omega_0/2,\omega_0\right)$,
where $\omega_0=\pi f_s$. The process is repeated in cascade for all the
levels.

The processing of $N$ levels requires to store additional informations: the
sum and the counter needed to evaluate the mean of 
$\left|c_n^k\right|^2$,
the current address inside the queue and the status of the queue.

The coefficients $c_n^k$ feed the register $\sum\left|c_n^k\right|^2$,
which stores the mean square amplitude of the high-pass filtered signal
representing the power spectrum $S\left(\omega\right)$
in the angular frequency range $\left(\omega_0/2^{k+1},\omega_0/2^k\right)$.

The status of the queue 
can be 0 (the queue has just been processed), 1 (the queue had one
input) or 2 (the queue is full and requires processing); it is incremented
by one each time an element is put in the queue and is reset to zero when it
is processed. At the initialization, the status is set to $-M+2$, so that
the queue is filled at least with $M$ elements before it is processed the
first time. 

As outlined above, Level 1 is updated only when two new samplings are 
available in Level 0. Therefore the update of Level 1 occurs with a frequency 
$f_s/2$,
and its processing with a frequency $f_s/4$, which corresponds to the
update frequency of Level 2, and in general Level $k$ is updated at a
frequency $f_s/2^k$ and processed at $f_s/2^{k+1}$. The sum of all
the processing frequencies for every queue, $\sum_n f_s/2^{k+1}$,
converges to $f_s$ as $N\to +\infty$. This means that the frequency at which
all the queues are processed is smaller than the sampling frequency $f_s$.
At each sampling time, our algorithm processes the lowest-level full queue,
if any.
This ensures that, by processing exactly one queue at each sampling time,
the output values $y_n^k$ never go to an already full queue. 
For the Daubechies wavelets with $M=4$ all the
operations can be performed in about $2\mathrm{\mu s}$.

Beyond the processing sofware on the DSP board two host programs run on the
host computer: the first uploads the DSP programs and runs it; the second
reads the memory of the DSP, without halting the CPU. This program is used
to get a real time display of the processed spectrum. The read program
accesses the DSP memory through the parallel port. Inside the DSK board,
data are read from memory by using a low priority DMA channel, and thus
without interfering with the CPU operations.

Another processing method, not yet implemented, involves the processing of
the cross spectrum of two signals. This is very useful to get rid of noise
sources such as afterpulsing from a photomultiplier detector.
This method involves the
parallel use of two processing structures like the one shown in Fig. 1. Each
detector signal feeds one of the TIMER input. The output of each structure
is a set of wavelet coefficients $c_{an}^k$
and $c_{bn}^k$. The averaged
product of the coefficients $\sum c_{an}^k c_{bn}^{k}$ gives the
desired cross spectrum. Due to the simultaneous operation of two processing
structures, the minimum sampling frequency is reduced by a factor of two.

\section{Experimental results}

To assess the performances of the Spectrum Analyzer we have performed test
measurements on a colloidal suspension and on a model system giving rise to a
power-law power spectrum.

The experimental setup is based on the traditional DLS one. A diagram of
the setup is shown in Fig. 2.
\begin{figure}
\begin{picture}(240,153)(0,87)
\thicklines
\put(205,216){LASER}
\path(202,213)(240,213)(240,227)(202,227)(202,213)
\path(202,220)(195,220)
\put(120,220){\circle{40}}
\put(133,198){C}
\put(132.5,220){\arc{47.0}{-0.72972766}{0.72972766}}
\put(167.5,220){\arc{47.0}{2.411865}{-2.411865}}
\put(153,198){F}
\put(163,221){Spatial}
\put(163,211){filter}
\path(160,205)(195,205)(195,235)(160,235)(160,205)
\thinlines
\path(160,218)(150,218)(90,222)
\path(160,222)(150,222)(90,218)
\thicklines
\path(24,100)(0,124)(20,144)(44,120)(24,100)
\put(10,120){PMT}
\thinlines
\path(35,135)(65,175)(120,220)(75,165)(35,135)
\path(31.571429,132.42857143)(35,135)(32.42857143,131.571429)
\thicklines
\path(33,137)(23,147)
\path(37,133)(47,123)
\put(46,126){D2}
\path(70,180)(63,187)
\path(80,170)(87,163)
\put(64,188){D1}
\put(56.3333,156.3333){\arc{48.4}{-1.4288993}{-0.14189705}}
\put(83.6666,183.6666){\arc{48.4}{1.7126934}{2.9996956}}
\put(52,170){L}
\thinlines
\put(64,140){\arc{84.852814}{1.5707963}{2.3561945}}
\thicklines
\path(64,87)(64,108)(102,108)(102,87)(64,87)
\put(73,99){ALV}
\put(66,89){PM-PD}
\thinlines
\path(102,97)(125,97)
\thicklines
\path(125,87)(125,108)(165,108)(165,87)(125,87)
\put(127,95){HM8035}
\thinlines
\path(165,97)(187,97)
\thicklines
\path(187,87)(187,108)(211,108)(211,87)(187,87)
\put(189,95){DSP}
\thinlines
\path(211,97)(220,97)
\thicklines
\path(220,87)(220,108)(240,108)(240,87)(220,87)
\put(223,95){PC}
\thinlines
\path(176,97)(176,127)(187,127)
\thicklines
\put(176,97){\circle*{2}}
\path(187,117)(187,138)(236,138)(236,117)(187,117)
\put(189,125){HM8021-3}
\thinlines
\path(113.5,97)(113.5,167)(125,167)
\thicklines
\put(113.5,97){\circle*{2}}
\path(125,157)(125,178)(170,178)(170,157)(125,157)
\put(127,165){ALV5000}
\thinlines
\path(170,167)(179,167)
\thicklines
\path(179,157)(179,178)(199,178)(199,157)(179,157)
\put(181,165){PC}
\end{picture}
\caption{Diagram of the experimental setup.}
\end{figure}
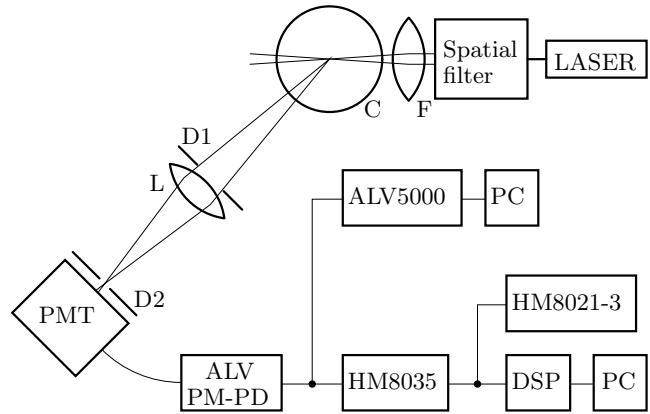
The beam coming from a JDS Uniphase 35mW CW
HeNe LASER model 1145P is spatially filtered and focused by the lens F 
at the center of the sample contained in
the cuvette C. Light scattered from the sample is collected by the lens L
and by the diaphragms D1 and D2. The scattered radiation eventually impinges
onto the photocathode of a EMI 9863 B04/350 photon counting Photomultiplier
Tube (PMT). The scattering angle can be changed by rotating the collecting
optics and the PMT around the cuvette by means of a goniometer. The current
pulses from the PMT are amplified and filtered by an ALV PM-PD
preamplifier-discriminator, which also digitizes the pulse train at TTL
levels. In the traditional DLS apparatus the signal is then fed to an
ALV5000 multi tau hardware correlator hosted in a personal computer.
Alternatively, the digitized signal at the output of the
amplifier-discriminator can be routed to a timer input of the DSP board,
which is connected to a personal computer for the configuration and
visualization of measurements. The DSP board is also connected to a HAMEG
1.6 GHz frequency counter model HM8021-3 and a HAMEG 20 MHz pulse generator
model HM8035 for testing purposes.

\subsection{Colloidal suspensions}

The test on the colloidal suspension were performed on calibrated
polystyrene latex of $0.115\mathrm{\mu m}$ diameter, diluted at a
volume fraction of about $\phi = 1.5\times 10^{-6}$.
For a dilute suspension of 
monodisperse brownian particles
the Homodyne time autocorrelation function of the intensity fluctuations in
the scattered light has the usual exponential form 
$C\left(t\right)= \left<I\right>^2
\left[1+A \exp \left(-\Gamma t\right)\right]$, 
where $\Gamma=2Dq^2$ and $D=k_BT/(6\pi\eta a)$ is the
Stokes-Einstein Diffusion coefficient, $\eta$ is the shear viscosity for the
solvent and $a$ the particle radius. The correlation function measured at a
scattering angle of $30^{\circ}$ is plotted in Fig. 3 as a function of the
lag time.
\begin{figure}
\includegraphics{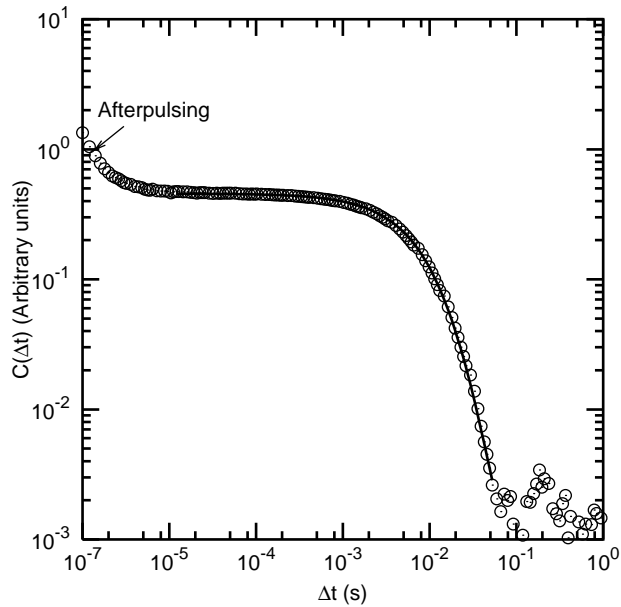}
\caption{Correlation function of the intensity fluctuations of light
scattered at $30^{\circ}$ by a colloidal suspension, in the Dynamic Light
Scattering regime, measured with ALV equipement.
Circles represent the experimental data. The continuous 
line represents a fit of the experimental data up to the second 
cumulant.}
\end{figure}
The experimental data show the
decay of the correlation function within a dynamic range of about $2.5$
decades. The run duration was $2700\mathrm{s}$.

At the longer delay times the correlation function is dominated by noise due
to the poor statistical sample accumulated. The additional contribution at
small lag times is due to the afterpulsing of the PMT. The afterpulses of
the PMT used to performed the measurement have a characteristic range of
correlation times between $0.1\mathrm{\mu s}$
and $3.2\mathrm{\mu s}$, which give rise to a stretched
exponential decay at small lags. The decay at lags larger than about  
$3.2\mathrm{\mu s}$ is determined by the intensity fluctuations due to
the brownian motion of the colloid. Data can be fit with an exponential
function $\exp\left(-\Gamma t\right)$. A more refined analisys of
the experimental data involves the inclusion of polidispersity effects in
the fitting procedure. This is traditionally performed by a cumulant analisys
of the correlation function \cite{pecora1}.
In the case of a small polidispersity
the correlation function up to the second cumulant is given by
\begin{eqnarray}
\label{eq_corr_diffusione}
C\left(t\right)\propto
\\* \nonumber
\exp \left(-\Gamma_0 \left|t\right|+\frac{1}{2}
\sigma^2t^2\right) \propto
\\* \nonumber
\int \exp
\left(-\Gamma \left|t\right|\right)\exp \left[-\frac{(\Gamma -\Gamma_0)^2}
{2\sigma^2}\right]\mathrm{d}\Gamma ,
\end{eqnarray}
where the second equality shows that polidispersity effects give rise to the
superposition of exponential decays centered around a central linewidth 
$\Gamma_0$, these decays being weighted by a gaussian function with
variance $\sigma$ representing the polidispersity. The best fit of the
experimental data with Eq. (\ref{eq_corr_diffusione})
is indicated by the solid line in Fig. 3 and
yields $\Gamma_0=140 \mathrm{Hz}$ and $\sigma = 42 \mathrm{Hz}$.

We performed the same experiment, by using the DSP WTSA and
the same homodyne detection scheme outlined above. According to the
Wiener-Kintchine theorem, the power spectrum of the fluctuations in the
intensity of light scattered by the colloidal suspension is the
Fourier transform of the correlation function $C\left(t\right)$. For diluted
monodisperse brownian particle the power spectrum has a Lorenzian shape: 
$S\left(\omega\right)\propto \Gamma/\left(\Gamma^2+\omega^2\right)$.

Figure 4 shows data for the measured power spectrum as a function of
angular frequency $\omega$.
\begin{figure}
\includegraphics{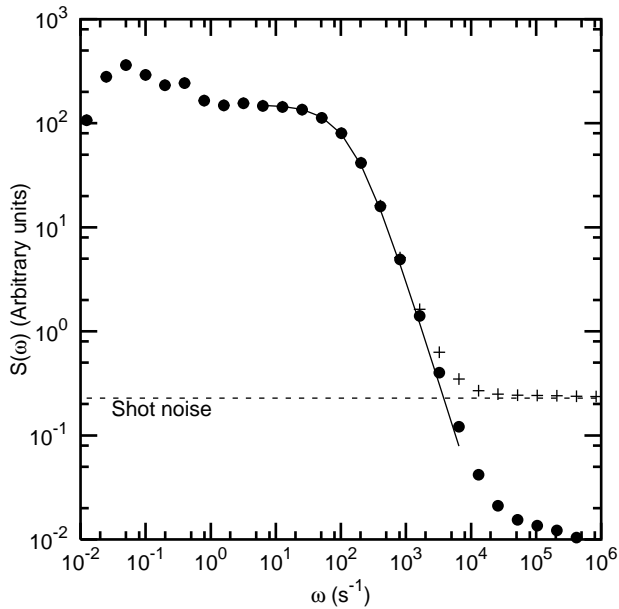}
\caption{Power spectrum of the intensity fluctuations of light
scattered at $30^{\circ}$ by a colloidal suspension, in the Dynamic Light
Scattering regime, measured with WTSA.
Crosses represent the raw data. The dashed line is the baseline due to the
shot noise. Circles represent the power spectrum
obtained by subtracting the baseline
from the raw data. The continuous line is the best fit of the experimental
data up to the second cumulant.}
\end{figure}
Raw data are represented with crosses. At the smaller
frequency the effect of noise due to the reduced statistical sample is
evident. This occurs at angular frequencies roughly smaller than 
$1\mathrm{rad/s}$,
corresponding to a time delay of the order of $6\mathrm{s}$.
However, this same noise affects the
correlation function shown in Fig 3 starting from about 
$t=10^{-1} \mathrm{s}$.
As the 
duration of the experiment is in both cases $2700\mathrm{s}$,
the wavelet processing
allows a better convergence at larger delay times: the timescale range
across which the power spectrum can be reliably determined is roughly two
decades more extended than that of the correlator. At larger frequency the
power spectrum in Fig. 4 falls down up to a constant base line, represented
by a horizontal dashed line. This baseline is determined by the shot noise.
In fact, it is well known that the detection of photons is a random
process characterized by a Poisson distribution and gives rise to a
delta-correlated white noise, whose power spectrum corresponds to the
average number of photons detected within the sampling time 
\cite{pecora1}.
\begin{equation}
C\left(t\right)\propto \frac{\left<I\left(t'\right)I\left(t'+t\right)\right>}
{\left<I^2\left(0\right)\right>}
-1+\left<n\right>\delta\left(t\right).
\end{equation}
In the frequency space this gives rise to a constant term $\left<n\right>$,
represented by the dashed line in Fig. 4, which adds
up to the power spectrum. The average number of photons $\left<n\right>$ can be
easily processed in real time and subtracted.

The circles in Fig. 4 show experimental data for the power spectrum after
the subtraction of the shot noise contribution. The data span a dynamic
range of more than $4$ decades. Data can be represented by a Lorenzian curve $\Gamma/\left(\omega^2+\Gamma^2\right)$.
In this case also, a more refined analisys
involves the inclusion of polidispersity effects. By taking the Fourier
transform of Eq. (\ref{eq_corr_diffusione}) we get:
\begin{equation}
\label{eq_spettro_diffusione}
S\left(\omega\right)
\propto \int \frac{\Gamma}{\omega^2+\Gamma^2}\exp 
\left[-\frac{\left(\Gamma -\Gamma_0\right)^2}
{2\sigma^2}\right]\mathrm{d}\Gamma .
\end{equation}
Therefore the power spectrum is the superposition of lorenzians weighted by
the same gaussian function introduced to describe the polidispersity in the
correlation analysis. We evaluated numerically the integral of
Eq. (\ref{eq_spettro_diffusione}), in order to 
fit the results in Fig. 4. However, each circle in Fig. 4
represents the integral of the power spectrum across an octave; 
this smooths even more the bell-shaped measured curve;
this effect has been considered in the fit procedure.
We obtained the best fit with $\Gamma_0=158 \mathrm{Hz}$ and
$\sigma=37 \mathrm{Hz}$, which
favourably compare with the results obtained with the correlator.
The best fitting curve is represented by the solid line. 

At the higher frequencies the spectrum is dominated by the contribution of
afterpulses to the intensity fluctuations which give rise to the tail in the
range $10^4-10^6 \mathrm{Hz}$ in Fig. 4.

The wavelet analysis of the data presents some advantages when compared to correlation techniques. A notable 
advantage is represented by the way background subtraction is achieved.  In correlation techniques a constant baseline 
has to be subtracted from the measured correlation function to recover data like those shown in Fig. 3. The baseline 
represents the correlation function at large la times, where correlations disappear. The intrinsic contrast of the correlation 
function, defined by the ratio between the amplitude of the completely  correlated part at $t=0$ and the uncorrelated 
part at large lag times, $\left[C\left(0\right)-C\left(+\infty\right)\right]/C\left(\infty\right)$,
cannot exceed the value of two.  Therefore, the signal and the background are of the same order of magnitude and 
the dynamic range is strongly limited by the accuracy in the baseline. The accuracy in the baseline can be determined 
from the arguments presented in Ref. \cite{degiorgio} and summarized below. Let us call $\tau_L$
the largest decay time in the correlation function and $T$ the duration of the run. Samples acquired at time lags smaller 
than $\tau_L$ are correlated, therefore the number of independent samples acquired during $T$ is 
$N_c=T/\tau_L$. This independent samples average out according to a normal distribution to give rise to the baseline.
Therefore, the accuracy in the baseline is of the order of $1/N_c^{1/2}$. In the case of the results presented in Fig. 3,
$\tau_L=40\mathrm{ms}$, $T=2700\mathrm{s}$, thus giving $1/N_c^{1/2}=4\times10^{-3}$,
which corresponds fairly well to the dynamic range of about $2.5$ decades of the measurement.
The same argument can be applied to the accuracy of the baseline in the power spectrum determined with the 
WTSA and presented in Fig. 4. In this case, however, the baseline is generated by the shot noise, which is correlated 
on a time of the order of the inverse of the sampling frequency $\omega_0$. The number $N$ of independent 
samples corresponds to the total number of samples accumulated at $\omega_0$. For the results presented in Fig. 4 
$N=10^9$ and the accuracy  of the baseline is of the order of $1/N^{1/2}=3\times 10^{-5}$.
By eliminating the contribution of afterpulsing to the spectrum as described in Sect. \ref{sec_programma},
the dynamic range could potentially be extended about five decades above the intrinsic contrast of the measurement,
which in this case amounts to about $10^3$.  It has to be remarked that in this case there''s no upper limit for the 
contrast, therefore a  dynamic range as high as $10^{10}$ could potentially be obtained.

\subsection{Model system}
\label{sistema_modello}

The ultimate task of the WTSA is the characterization of signals
characterized by a wide distribution of timescales.

To assess its performances we have chosen a simple model system made up by
strongly diluted diffusing brownian particles illuminated by a wide
collimated laser beam. This system gives rise to a nice power-law power
spectrum with the exponent $-3/2$. A detailed derivation of the predicted power
spectrum is presented in Appendix. The sample is a colloidal suspension of
polystyrene latex spheres of $10\mathrm{\mu m}$
diameter, suspended in water at a
concentration of roughly $10^4\mathrm{particles/cm}^3$. The solvent is a
mixture of equal volumes of water and deuterated water, so that its density
is matched to that of the colloid within 0.1\%. The sample is illuminated by
a gaussian laser beam with a diameter at $1/e^2$ of about $1\mathrm{mm}$. The
scattering volume is delimited by two slits perpendicular to the beam.
Therefore the scattering volume is determined in one direction by the hard
edge of the slits and in the other direction by the gaussian profile of the
beam. The scattered light is collected at $90^{\circ}$
with respect to the main beam.

Due to the strongly diluted sample and to the wide beam illumination,
fluctuations in the intensity of the scattered light are
mostly determined by
changes in the number of particles in the scattering volume.
By using a wide laser beam,
the coherence areas onto the photocatode of the PMT are very
small. Therefore, the photocatode collects the averaged contribution coming
from many coherence areas. In this way the component of the correlation
function due to the diffusion of particles has a vanishingly small contrast,
when compared to the averaged intensity measured by the photodetector. The
signal onto the PMT thereby represents the superposition of the intensities
scattered by single particles. As it is shown in the Appendix, intensity
fluctuations spanning a wide frequency range are excited when particles
cross the hard edge at the boundary of the scattering volume.

Figure 5 shows the correlation function of the fluctuations in the scattered
intensity.
\begin{figure}
\includegraphics{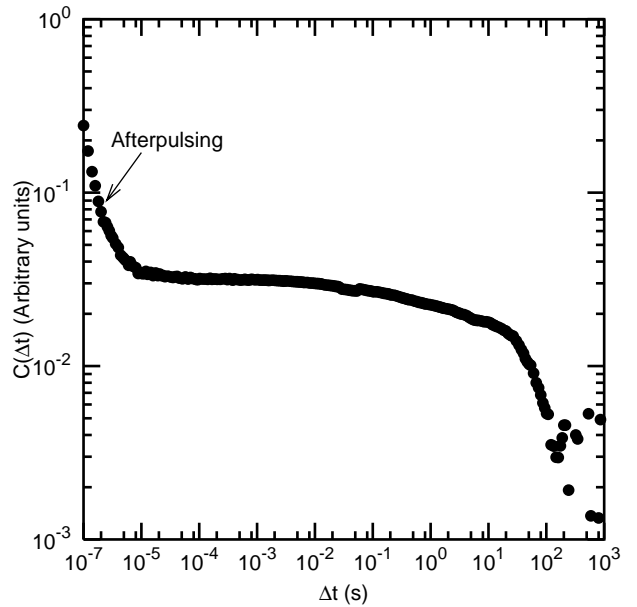}
\caption{Correlation function of the intensity fluctuations 
in the model system, measured with ALV equipement.}
\end{figure}
The correlation function shows an almost featureless
structure and a narrow dynamic range. The correlation function is almost flat
within about six decades in lag time, due to the long tails associated with
the slowest modes. The peak at small lags still represents the contribution
of afterpulses. The correlation function eventually decays at times of the
order of $60\mathrm{s}$. However, this decay is an artifact due to the finite 
measurement time.  In fact,
correlation analysis assumes that the correlation function decays to 
zero at large delays, once its baseline has been subtracted. The baseline is 
usually evaluated from the value of the correlation function at very large delays. 
However, for time-scale invariant processes, the baseline still contains the 
contribution of the slowest modes. Therefore, an increase of the acquisition time 
involves a decay at a larger time. As a rule of thumb, experimental data for the 
correlation function are reliable up to delays roughly 100 times smaller than 
the acquisition time. This is a feature of correlation techniques which strongly 
limits their ability in characterizing processes where an upper time scale is 
absent.

Figure 6 shows the power spectrum of the fluctuations measured by means of
the WTSA.
\begin{figure}
\includegraphics{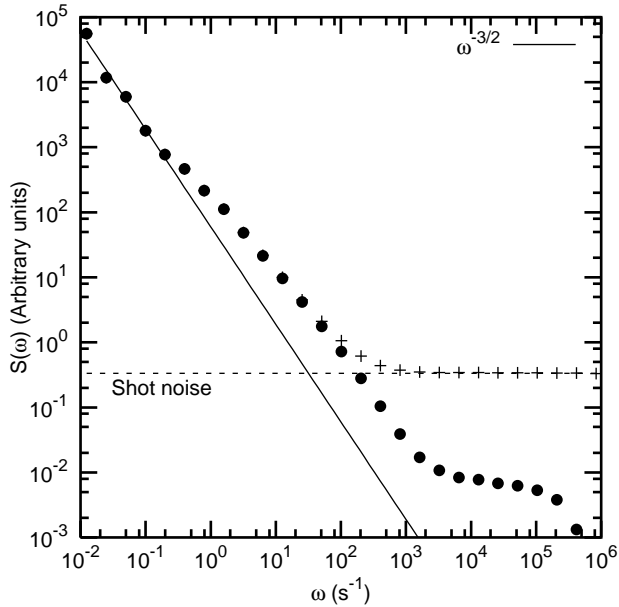}
\caption{Power spectrum of the intensity fluctuations 
in the model system, measured with WTSA.
Crosses represent the raw data. The dashed line is the baseline due to the
shot noise. Circles represent the power spectrum
obtained by subtracting the baseline
from the raw data. The continuous line represents a power-law with exponent
$-3/2$ for comparison.}
\end{figure}
The crosses represent the raw data, the
horizontal line the background level due to shot noise and the circles the
subtracted data representing the power spectrum. The most prominent feature of
the results in Fig. 6 is the impressive dynamic range spanning about 8 decades.
The measured power spectrum
decays approximately as a power law with an exponent of the order of $-1.2$.
The straight line represents a power law with exponent $-3/2$ for
comparison. At the larger frequency, the contribution of afterpulses is
apparent.

Deviations from the power law behavior at intermediate frequencies can be
attributed to a residual contribution of the diffusion of particles to the 
correlation function. Such a contribution cannot be completely eliminated,
even by collecting a large number of coherence areas onto the multiplier. As
outlined above, the diffusion of particles gives rise to fluctuations in the
scattered intensity at $90^{\circ}$, characterized by lorentzian spectrum with 
a linewidth of the order of $\Gamma=20\mathrm{Hz}$.
Deviations form the power-law behavior occur at frequencies roughly centered 
around $\Gamma$. 
However, a more refined analysis of the data, including the evaluation of
contributions due to diffusion and due to number fluctuations to the spectrum,
is beyond the aim of this paper.

The wavelet transform analysis is particularly effective for processes where the power law exponent $\alpha$ is larger
than $1$, like the model system just described. In fact, correlation techniques completely fail when applied to such 
systems. For exponents smaller than one the spectral behavior can be characterized either by measuring the power 
spectrum $S\left(\omega\right) \propto 1/\omega^{\alpha}$ or by measuring the correlation function, which also 
exhibits a power-law behavior $C\left(t\right) \propto t^{1-\alpha}$. In the limit $\alpha \to1$ the correlation function 
becomes flat. For $\alpha > 1$ the correlation function is still flat and time independent. Naively speaking, this is due 
to the strong divergence of the power spectrum at small frequency which behaves as a delta shaped spectrum. The 
diverging energy of the small frequency modes  gets uniformly distributed along the lag time axis of the correlation 
function, which becomes flat. More rigorously, the correlation function can be obtained by Fourier transforming the 
power-law spectrum, regularized to avoid its divergence at $\omega = 0$: 
\begin{equation}
C\left(t\right) \propto  
\int {
\lim_{\epsilon \to 0} \frac{1}{\epsilon+\omega^{\alpha}} \exp{\left(-i\omega t\right)} \mathrm{d}\omega
}.
\end{equation}
For $\alpha>1$, the spectrum has an integrable power law tail for $\omega \to +\infty$. It approximates a delta 
function $\delta\left(\omega\right) $ as $\epsilon \to 0$. Therefore, for $\alpha \ge 1$, $C\left(t\right)$
is a constant: a correlation function like that shown in Fig. 3 is completely featureless and useless.

An important application of the  WTSA is the time dependent analisys of transient behavior during processes such 
as gelation \cite{kroon, wegdam} and colloidal aggregation \cite{vailati93}. In general, during such 
transients the highest frequencies are excited at first, and lower frequencies get gradually excited as time goes by. This 
introduces a  characteristic lag time $\tau_c$ marking the boundary between  excited and non-excited modes. 
The proper characterization of this processes involves the measurement of the power spectrum (or correlation function)
from data sampled along a duration time $T$ larger then $\tau_c$, so to be able to resolve $\tau_c$. In principle a 
longer $T$ allows to measure the power spectrum or correlation function with a higher accuracy, due to the increase of 
the number of independent sample processed. However the lag time $\tau_c$ changes during the transient. Therefore, 
a compromise has to be reached between the need to keep $T$ not much larger than $\tau_c$ and the desire to measure 
the spectrum with a high accuracy.

As far as the WTSA is concerned, the same argument used to estimate the accuracy of the baseline can be used to 
get an order of magnitude estimate of the number of samplings needed to get a chosen accuracy. Suppose that we 
want to obtain the power spectrum at a frequency $\omega_k$ with an accuracy better than $a$. As it was shown in 
section \ref{sezione_wavelet_analysis}, the power spectrum is calculated by averaging the squared wavelet coefficients:
$S\left(\omega_k\right) \propto \sum_n \left|c^k_n\right|^2$.
We recall here that the coefficients $c^k_n$ represent the wavelet coefficients evaluated at subsequent times $n$.
Although this argument is not rigorous, for a generic random signal, it is reasonable to assume that the coefficients 
evaluated at different times are independent. Therefore the percentual error on 
$S\left(\omega_k\right)$ scales as $1/N_k^{1/2}$, where $N_k$ is the number of samplings of the signal accumulated 
at the frequency $\omega_k$. In this way, a $1\%$ in $S\left(\omega_k\right)$ is achieved by sampling the signal up 
to frequencies of the order of $10000\omega_k$. However, an order of magnitude estimate of the power spectrum can 
be obtained from a few samplings at $\omega_k$. 

Therefore, a very important feature of the wavelet analysis, when compared to correlation
techniques, is its ability to characterize the power spectrum at small
frequencies, without introducing artifacts related to the finite acquisition
time. This can be appreciated from Fig. 6, where the smallest experimental
angular frequency is of the order of $10^{-2}\mathrm{Hz}$, roughly corresponding
to the acquisition time of $2700\mathrm{s}$. The wavelet analysis only yields
an order of magnitude estimate of the amplitude of this mode, due to the absence
of a significative statistical sample. However, the wide dynamic range involved
in the power spectrum makes it almost insensitive to background subtraction at
small frequencies, so that no characteristic times associate with the duration 
of the experiment are introduced in the subtraction. Therefore, when compared with
correlation techniques, wavelet analysis allows a dramatic increase of the 
investigated frequency range, this increase amounting to almost two decades. This
feature makes the WTSA the ideal tool to investigate slow processes, 
non-stationary processes, time-scale invariant processes and processes 
characterized by a wide distribution of time-scales in general.

\acknowledgements

We thank Texas Instruments DSP University program for the gift of the DSK board
and for technical support. We also thank M. Giglio for discussion
and support.

\appendix

\section{Power spectrum of the number fluctuations of brownian particles in a given volume.}

In this appendix we will derive the power spectrum determined by the number
fluctuations of brownian particles diffusing within a scattering volume
delimited by sharp edges on two sides and by the gaussian profile of a beam
on the two perpendicular sides. We will assume that the sample is illuminated
by a wide collimated beam, so that spatial coherence effects can be neglected.
Some of the arguments we use in the following are also used in the derivation
of fluorescence correlation spectroscopy. \cite{pecora1}

Given $N$ particles, at position $\vec{x}_n$, the measured intensity $I_t$
at time $t$ is the sum of the intensities scattered by each particle,
since light is collected over several speckles, and interference plays no
role. Moreover, the intensity scattered by each particle is proportional to
the intensity $I\left( \vec{x}\right) $ of the main beam at the position of
the particle: 
\begin{equation}
I_t\propto \sum_{n=0}^{N-1}{I\left[ \vec{x}_n\left(t\right)\right]}.
\end{equation}
Roughly speaking, $I_t$ is proportional to the number of particles inside
the main beam, weighted by the intensity in that point. The temporal
correlation function is thus: 
\begin{equation}
C_I\left(\tau\right) =\left< I_t I_{t+\tau}\right> \propto
\sum_{n,m=0}^{N-1}\left< I\left[\vec{x}_n\left(t\right)\right] 
I\left[ \vec{x}_m\left(t+\tau\right) \right] \right> , 
\end{equation}
where the brackets represent the average over $t$. By
assuming that the particles are non-interacting, their positions are not
correlated:
\begin{equation}
C_I\left(\tau\right) \propto N\left(N-1\right)
\bar{I}^2+N\left<I\left[\vec{x}\left(t\right)\right]
I\left[\vec{x}\left(t+\tau\right)\right]\right> ,
\label{rumore_conteggio_corr_con_costante}
\end{equation}
where the bar represents the average over the volume $V$ of the whole
cell. From this equation, we derive the relative mean square value of the
fluctuations of $I_{t}$: 
\begin{equation}
\frac{\left<I_t^2\right>-\left<I_t\right>^2}
{\left<I_t\right>^2}=\frac{C_I\left(0\right)
-C_I\left(+\infty\right)}{C_I\left(+\infty\right)}=\frac{1}{N}
\frac{\bar{I^2}-\bar{I}^2}{\bar{I}^2} .
\end{equation}
Roughly speaking, $\left(\bar{I^2}-\bar{I}^2\right) /\bar{I}^2$ is
the ratio $V/V_b$, where $V_b$ is the volume inside the main beam. This
statement can be easily proved for an hypotetic beam with constant intensity
inside $V_b$, vanishing outside. By calling $N_b$ the number of
particles inside the beam: 
\begin{equation}
\frac{\left<I_t^2\right>-\left<I_t\right>^2}
{\left<I_t\right>^2}=\frac{1}{N_b} .
\end{equation}
This equation is analogous to the relation between number fluctuation and
mean number for Poisson distribution. In order for the fluctuations to be
strong, only a few particles must be inside the beam simultaneously.

The diffusion of each particle can be described as a random walk; 
$P_{\tau}\left(\vec{y}-\vec{x}\right)$, the probability that a particle in 
$\vec{x}$ goes in $\vec{y}$ in a time $\tau$, is thus: 
\begin{equation}
P_{\tau}\left(\Delta \vec{x}\right)= \frac{1}{\left[\sqrt{2\pi}%
\sigma\left(\tau\right)\right]^3} e^{\displaystyle -\frac{\Delta \vec{x}^2}{%
2\sigma^2\left(\tau\right)} } ,
\end{equation}
where $\sigma\left(\tau\right)=\sqrt{D\left|\tau\right|}$ and $D$ is the
diffusion coefficient. By using this probability distribution, we can
explicitly rewrite Eq. (\ref{rumore_conteggio_corr_con_costante}): 
\begin{equation}
C_I\left(\tau\right) \propto c+ \int{I\left(\vec{x}\right) I\left(\vec{y}\right)
P_{\tau}\left(\vec{y}-\vec{x}\right) \mathrm{d}\vec{x}\mathrm{d}
\vec{y} } ,
\end{equation}
where $c$ is a constant.
The convolution operator is diagonalized by Fourier transform: 
\begin{equation}
\label{eq_C_I_tau}
C_I\left(\tau\right) = c+ \int{\left|I\left(\vec{q}\right)\right|^2
P_{\tau}\left(\vec{q}\right) \mathrm{d}\vec{q} },
\end{equation}
where $P_{\tau}\left(\vec{q}\right)$ and $I\left(\vec{q}\right)$ are the
Fourier transforms in $\vec{x}$ of $P_{\tau}\left(\vec{x}\right)$ and 
$I\left(\vec{x}\right)$: 
\begin{equation}
\label{eq_P_tau_q}
P_{\tau}\left(\vec{q}\right) = e^{\displaystyle -
\frac{1}{2}q^2\sigma^2\left(\tau\right) }.
\end{equation}
By Fourier transforming Eq. (\ref{eq_C_I_tau}) in $\tau$: 
\begin{equation}
\label{eq_corr_conteggio_spettro_generico}
S_I\left(\omega\right) = \int{\left|I\left(\vec{q}\right)\right|^2
P_{\omega}\left(\vec{q}\right) \mathrm{d}\vec{q} } ,
\end{equation}
where $P_{\omega}\left(\vec{q}\right)$ is the Fourier transform in $\tau$ of 
$P_{\tau}\left(\vec{q}\right)$ as defined by Eq. (\ref{eq_P_tau_q}):
\begin{equation}
P_{\omega}\left(\vec{q}\right) = \frac{1}{\pi} \frac{Dq^2}{\omega^2+D^2q^4} .
\end{equation}

Now, we explicitly give the intensity distribution $I\left(\vec{x}\right)$
within the scattering volume:
\begin{equation}
I\left(x,y,z\right)=I_0\chi_{\left[-L/2,L/2\right]}
\left(z\right)
e^{\displaystyle-\frac{x^2+y^2}{2R^2}},
\end{equation}
where $L$ is the length of the portion of the beam we observe, $R$ is the
radius of the main beam at $e^{-1/2}$ and $I_{0}$ is the intensity at the
center of the main beam. The Fourier transform is:
\begin{equation}
\label{eq_I_q}
I\left(q_x,q_y,q_z\right)=\frac{4\pi I_0}{Q_LQ_R^2}
\frac{\sin\left(q_z/Q_L\right)}{q_z/Q_L}e^{\displaystyle-
\frac{q_x^2+q_y^2}{2Q_R^2}} ,
\end{equation}
where $Q_L=2/L$ e $Q_R=1/R$. By inserting the explicit expressions 
of Eq. (\ref{eq_P_tau_q}) and Eq. (\ref{eq_I_q})
in Eq. (\ref{eq_corr_conteggio_spettro_generico}): 
\begin{eqnarray}
S_I\left(\omega\right)=\frac{16\pi I_0^2}{Q_L^2Q_R^4}
\int {\frac{\sin^2\left(q_z/Q_L\right)}{\left(q_z/Q_L\right)^2}
e^{\displaystyle-\frac{q_x^2+q_y^2}{Q_R^2}}
}\times
\\* \nonumber
\frac{D\left(
q_x^2+q_y^2+q_z^2\right)}{\omega^2+D^2\left(
q_x^2+q_y^2+q_z^2\right)^2}\mathrm{d}q_x\mathrm{d}q_y
\mathrm{d}q_z .
\end{eqnarray}

Now, we impose that $\omega \gg DQ_R^2$ and $\omega \gg DQ_L^2$. This means
that $\omega$ corresponds to times much shorter than the time needed by a
particle to travel across the main beam. Under this condition,
 we can
neglect $q_x$ and $q_y$ at the denominator, since the gaussian
part in the integrand imposes $\left|q_x\right| \lesssim Q_R$ and
$\left|q_y\right| \lesssim Q_R$; then, we integrate 
over $q_x$ and $q_y$:
\begin{equation}
S_I\left(\omega\right) = \frac{16\pi I_0^2}{Q_L^2 Q_R^2} \int{\frac{\sin^2\left(q_z/Q_L\right)}{\left(q_z/Q_L\right)^2} 
\frac{
D\left(q_z^2+2Q_R^2\right)}{\omega^2+D^2 q_z^4}\mathrm{d}q_z } .
\end{equation}
The $\sin^2\left(q_z/Q_L\right)$ has fast oscillations, and 
averages to $1/2$ for $q_z \gg Q_L$: 
\begin{equation}
S_I\left(\omega\right) = \frac{8\pi I_0^2}{Q_R^2}
\int{\frac{Q_L^2}{q_z^2+Q_L^2} \frac{D\left(q_z^2+2Q_R^2\right)}
{\omega^2+D^2 q_z^4}\mathrm{d} q_z } .
\end{equation}
The leading term in the integral is the $q_z^{-2}$ coming 
from the sinc function.
This means that the behaviour of the measured intensity is mainly due to
particles crossing the sharp edges of the slits, moving along the main beam,
while the movements inside and outside the beam can be neglected.

Since $Q_{R}$ and $Q_{L}$ are of the same order of magnitude: 
\begin{equation}
S_I\left(\omega\right)=8\pi I_0^2D\int {\frac{1}{\omega^2
+D^2 q_z^4}\mathrm{d}q_z} .
\end{equation}
The integral converges; by substituting $q_z=x\sqrt{\omega}$: 
\begin{equation}
S_I\left(\omega\right) \propto \omega^{-3/2}.
\end{equation}

\end{document}